# Efficient algorithms for scheduling equal-length jobs with processing set restrictions on uniform parallel batch machines

Shuguang Li[*,1,2]

[1] Key Laboratory of Intelligent Information Processing in Universities of Shandong (Shandong Institute of Business and Technology), Yantai 264005, China;
[2] College of Computer Science and Technology, Shandong Institute of Business and Technology, Yantai 264005, China
*Address correspondence to this author at College of Computer Science and Technology, Shandong Institute of Business and Technology, 191 Binhai Middle Road, Laishan District, Yantai, Shandong Province, China. Postcard: 264005; E-mail: sgliytu@hotmail.com

**Abstract**: We consider the problem of scheduling jobs with equal lengths on uniform parallel batch machines with non-identical capacities where each job can only be processed on a specified subset of machines called its processing set. For the case of equal release times, we give efficient exact algorithms for various objective functions. For the case of unequal release times, we give efficient exact algorithms for minimizing makespan.

## 1. Introduction

The problem of scheduling uniform parallel batch machines with processing set restrictions can be defined as follows. Let $\mathcal{J} = \{1, 2, \ldots, n\}$ be a set of jobs and $\mathcal{M} = \{M_1, M_2, \ldots, M_m\}$ be a set of uniform parallel batch machines. Job $j$ ($j = 1, 2, \ldots, n$) becomes available at its *release time* $r_j \geq 0$ and requires $p_j \geq 0$ units of processing called its *length*. For each job $j$, let $\mathcal{M}_j \subseteq \mathcal{M}$ be the set of the *eligible machines* which are capable of processing the job, called its *processing set*. Each job is assigned to exactly one machine, i.e., job preemption is not allowed. Machine $M_i$ ($i = 1, 2, \ldots, m$) has a *speed* $v_i \geq 1$ and a *capacity* $K_i < n$. The impact of the speed is that $M_i$ can carry out $v_i$ units of processing in one time unit. That is, if job $j$ is assigned to machine $M_i$, then it requires $p_j / v_i$ *processing time* to be completed. Machine $M_i$ can process several jobs as a batch simultaneously as long as the total number of these jobs does not exceed $K_i$. The *length* of a batch is the maximum of the lengths of the jobs belonging to it. Jobs in the same batch have a common start time and a common completion time. The goal is to schedule the jobs on the machines in a manner that optimizes some objective function.

Two classes of objectives are considered: the min-sum objective and the min-max objective. Specifically, let $f_j : [0, +\infty) \to [0, +\infty)$ ($j = 1, 2, \ldots, n$) be a non-decreasing function. Additional parameters that may be included for job $j$ are its *due date* $d_j$ and its weight $w_j$. For a particular schedule $\Sigma$, let $C_j(\Sigma)$ denote the *completion time* of job $j$ in $\Sigma$. Let $T_j(\Sigma) = \max\{C_j(\Sigma) - d_j, 0\}$ denote the *tardiness* of job $j$ in $\Sigma$. Let $U_j(\Sigma) = 1$ if $C_j(\Sigma) > d_j$ and

$U_j(\Sigma) = 0$ otherwise. (In the rest of this paper, we safely ignore $\Sigma$ in the notations without causing confusion.) The objectives of minimizing $\sum_{j=1}^{n} f_j(T_j)$ and $\max_{j=1,2,\ldots,n}\{f_j(T_j)\}$ will be considered. Following [1, 2], the models can be denoted as $Q\,|\,r_j, d_j, \mathcal{M}_j, p-batch, K_i\,|\sum f_j(T_j)$ and $Q\,|\,r_j, d_j, \mathcal{M}_j, p-batch, K_i\,|\max\{f_j(T_j)\}$.

Many popular scheduling objectives are covered by the two models, such as total weighted completion time ($\sum w_j C_j$) minimization, total weighted tardiness ($\sum w_j T_j$) minimization, weighted number of tardy jobs ($\sum w_j U_j$) minimization, makespan ($C_{\max} = \max C_j$) minimization, and maximum weighted tardiness ($\max\{w_j(T_j)\}$) minimization. Most of such problems are NP-hard even for the special cases where all $v_i = 1$, all $K_i = 1$ and all $\mathcal{M}_j = \mathcal{M}$ [3, 4, 5]. Thus, we are interested in polynomial time exact algorithms for some important special cases of the problems [6].

In this paper, we focus on an important special case where all jobs have equal lengths. The problems under study can be denoted as $Q\,|\,r_j, p_j = p, d_j, \mathcal{M}_j, p-batch, K_i\,|\sum f_j(T_j)$ and $Q\,|\,r_j, p_j = p, d_j, \mathcal{M}_j, p-batch, K_i\,|\max\{f_j(T_j)\}$. Li [7] presented polynomial time algorithms for uniform parallel machines scheduling problems $Q\,|\,p_j = p, d_j, \mathcal{M}_j\,|\sum f_j(T_j)$ and $Q\,|\,p_j = p, d_j, \mathcal{M}_j\,|\max\{f_j(T_j)\}$. (the special cases of $Q\,|\,r_j, p_j = p, d_j, \mathcal{M}_j, p-batch, K_i\,|\sum f_j(T_j)$ and $Q\,|\,r_j, p_j = p, d_j, \mathcal{M}_j, p-batch, K_i\,|\max\{f_j(T_j)\}$ where all $r_j = 0$ and all $K_i = 1$.) We extend the results obtained in [7] to uniform parallel batch machines. Moreover, for minimizing makespan, we allow unequal release times and get an algorithm for arbitrary processing set restrictions.

Leung and Li [8] discussed several special cases of processing set restrictions. The processing sets of the jobs are *inclusive*, if for any two jobs $j_1$ and $j_2$, either $\mathcal{M}_{j_1} \subseteq \mathcal{M}_{j_2}$, or $\mathcal{M}_{j_2} \subseteq \mathcal{M}_{j_1}$. The processing sets are *nested*, if for any two jobs $j_1$ and $j_2$, either $\mathcal{M}_{j_1} \cap \mathcal{M}_{j_2} = \emptyset$, or $\mathcal{M}_{j_1} \subseteq \mathcal{M}_{j_2}$, or $\mathcal{M}_{j_2} \subseteq \mathcal{M}_{j_1}$. The processing sets are *interval*, if for any job $j$, $\mathcal{M}_j = \{M_{a_j}, M_{a_j+1}, \ldots, M_{b_j}\}$ for some $1 \leq a_j \leq b_j \leq m$. The processing sets are *tree-hierarchical*, if each machine is represented by a tree node, and each job $j$ is associated with a tree node $M_{a_j}$, such that $\mathcal{M}_j$ is exactly the set of the machines consisting of all the nodes on the unique path from $M_{a_j}$ to the root of the tree.

The remainder of this paper is organized as follows. In Section 2, the related researches are reviewed. In Section 3, we consider the case of equal release times. We present an algorithm with running time $O(n^3 m + n^2 m \log(mn))$ for $Q\,|\,p_j = p, d_j, \mathcal{M}_j, p-batch, K_i\,|\sum f_j(T_j)$ (the special case of $Q\,|\,r_j, p_j = p, d_j, \mathcal{M}_j, p-batch, K_i\,|\sum f_j(T_j)$ where all $r_j = 0$), as well as an algorithm with running time $O(n^{5/2} m^{3/2} \log(mn))$ for $Q\,|\,p_j = p, d_j, \mathcal{M}_j, p-batch, K_i\,|\max\{f_j(T_j)\}$ (the special case of $Q\,|\,r_j, p_j = p, d_j, \mathcal{M}_j, p-batch, K_i\,|\max\{f_j(T_j)\}$ where all $r_j = 0$). In Section 4, we consider the case of unequal release times. We present an algorithm with running time $O(n^{5/2} m^{3/2} \log(mn))$ for $Q\,|\,r_j, p_j = p, \mathcal{M}_j, p-batch, K_i\,|\,C_{\max}$. Section 5 presents the conclusions and future directions of research.

2. **Literature Review**

The problem studied in this paper combines two important sub-fields of scheduling theory: scheduling with processing set restrictions and parallel batch scheduling. The two sub-fields have received intense study in the literature, see the survey papers [8] and [9-11] respectively. However, only several papers combined the two-subfields into a unified framework [12-18]. In the problems studied in these papers except the last two, each job has a size and a machine can process several jobs simultaneously as a batch as long as the total size of these jobs does not exceed its capacity. Any machine cannot process the jobs whose sizes are larger than its capacity. Thus, for each job, the machines whose capacities are not less than its size form its processing set. Clearly, the processing sets of the jobs are inclusive. In [17], Li presented two algorithms with approximation ratios 3 and 9/4 for the problem of minimizing makespan on parallel batch machines with inclusive processing set restrictions, where the jobs have arbitrary lengths and the machines have the same speed. In [18], Li studied parallel batch scheduling with nest processing set restrictions to minimize makespan, and presented a $(3-1/m)$-approximation algorithm for the case of equal release times and a polynomial time approximation scheme (PTAS) for the case of unequal release times.

For scheduling with processing set restrictions, the review focuses on the case of equal job lengths. Lin and Li [19] obtained an algorithm for $P\,|\,p_j=p,\mathcal{M}_j\,|\,C_{max}$ (the special case of $Q\,|\,r_j,p_j=p,\mathcal{M}_j\,|\,C_{max}$ where all $r_j=0$ and all $v_i=1$) that runs in $O(n^3 \log n)$ time, and generalized the algorithm to solve $Q\,|\,p_j=p,\mathcal{M}_j\,|\,C_{max}$ in $O(n^3 \log(nv_{lcm}))$ time, where $v_{lcm}$ denotes the least common multiple of $v_1, v_2, \ldots, v_m$. They also obtained an algorithm for $P\,|\,p_j=p,\mathcal{M}_j(\text{interval})\,|\,C_{max}$ (the special case of $P\,|\,p_j=p,\mathcal{M}_j\,|\,C_{max}$ with interval processing set restrictions) that runs in $O(m^2+mn)$ time. Harvey et al. [20] independently developed an algorithm for $P\,|\,p_j=p,\mathcal{M}_j\,|\,C_{max}$ that runs in $O(n^2 m)$ time. Brucker et al. [21] presented algorithms running in $O(n^2 m(n+\log m))$ time for $Q\,|\,p_j=p,d_j,\mathcal{M}_j\,|\,\sum w_j T_j$, $Q\,|\,p_j=p,d_j,\mathcal{M}_j\,|\,\sum w_j U_j$, $P\,|\,r_j,p_j=1,d_j,\mathcal{M}_j\,|\,\sum w_j T_j$ and $P\,|\,r_j,p_j=1,d_j,\mathcal{M}_j\,|\,\sum w_j U_j$. Li [7] presented an algorithm for $Q\,|\,p_j=p,d_j,\mathcal{M}_j\,|\,\sum f_j(T_j)$ that runs in $O(n^3 m+n^2 m \log(mn))$ time. For the special cases where $f_j(T_j)=C_j$ or $f_j(T_j)=U_j$, the running time of the algorithm can be improved to $O(n^{5/2} m \log n)$. He also presented an algorithm for $Q\,|\,p_j=p,d_j,\mathcal{M}_j\,|\,\max\{f_j(T_j)\}$ that runs in $O(n^{5/2} m \log(mn))$ time. For the special cases where $f_j(T_j)=C_j$ (i.e., $Q\,|\,p_j=p,\mathcal{M}_j\,|\,C_{max}$), the running time of the algorithm can be improved to $O(n^2(m+\log(nv_{max}))\log n)$, where $v_{max}=\max\{v_j\}$. Lee et al. [22] showed that $Q\,|\,r_j,p_j=p,\mathcal{M}_j\,|\,C_{max}$ can be solved in $O(m^{3/2} n^{5/2} \log(mn))$ time, and $P\,|\,r_j,p_j=p,\mathcal{M}_j\,|\,C_{max}$ (the special case of $Q\,|\,r_j,p_j=p,\mathcal{M}_j\,|\,C_{max}$ where all $v_i=1$) can be solved in $O(m^{3/2} n^{5/2} \log n)$ time. Shabtay et al. [23] obtained various results for several problems of scheduling uniform machines with equal length jobs, processing set restrictions and job rejection.

Pinedo [24] and Glass and Mills [25] presented algorithms for $P\,|\,p_j=p,\mathcal{M}_j(\text{nested})\,|\,C_{max}$ (the special case of $P\,|\,p_j=p,\mathcal{M}_j\,|\,C_{max}$ with nested processing set restrictions) that run in time $O(n \log n)$ and $O(m^2)$ time respectively. Li and Li [26] presented algorithms for $P\,|\,r_j,p_j=p,\mathcal{M}_j(\text{inclusive})\,|\,C_{max}$ and $P\,|\,r_j,p_j=p,\mathcal{M}_j(\text{tree})\,|\,C_{max}$ (the special cases of $P\,|\,r_j,p_j=p,\mathcal{M}_j\,|\,C_{max}$ with inclusive and tree-hierarchical processing set restrictions) that run in $O(n^2+mn \log n)$ time. For uniform machines, they showed that $Q\,|\,r_j,p_j=p,\mathcal{M}_j(\text{inclusive})\,|\,C_{max}$ and $Q\,|\,r_j,p_j=p,\mathcal{M}_j(\text{tree})\,|\,C_{max}$ can be solved in $O(mn^2 \log m)$ time. Later, Li and Lee [27] developed an improved algorithm for $P|r_j, p_j = p, \mathcal{M}_j(\text{inclusive})|C_{\max}$ that runs in $O(\min\{m, \log n\}n \log n)$ time, and an improved algorithm for $P|r_j, p_j = p, \mathcal{M}_j(\text{tree})|C_{\max}$ that runs in $O(mn \log n)$ time.

For parallel batch scheduling, the review focuses on equal job lengths or uniform parallel batch machines. Liu et al. [28] presented an algorithm for $P|r_j, p_j = p, p-batch, B|C_{\max}$ (the special case of $Q|r_j, p_j = p, \mathcal{M}_j, p-batch, K_i|C_{\max}$ where all $\mathcal{M}_j = \mathcal{M}$, all $K_i = B$ and all $v_i = 1$) that runs in $O(n \log n)$ time. Ozturk et al. [29] presented a 2-approximation algorithm for the problem of scheduling jobs with equal lengths, unequal release times and sizes on identical parallel batch machines (all $K_i = B$) to minimize makespan. Wang and Leung [13] studied the problem of scheduling jobs with equal lengths and arbitrary sizes on parallel batch machines (all $v_i = 1$) with non-identical capacities. The problem fits into the model of scheduling with inclusive processing set restrictions, since each job can only be processed by the machines whose capacities are not less than the size of the job. Wang and Leung presented a 2-approximation algorithm, as well as an algorithm with asymptotic approximation ratio 3/2 for the problem. Li et al. [30] proposed several heuristics for the problem of scheduling jobs with unequal lengths, release times and sizes on uniform parallel batch machines with identical capacities to minimize makespan. Zhou et al. [31] presented an effective discrete differential evolution algorithm for the problem of scheduling jobs with unequal lengths and sizes on uniform parallel batch machines with non-identical capacities to minimize makespan. Both [30] and [31] have not included processing set restrictions.

## 3. Equal release times

In this section, we consider the case of equal release times. We will present algorithms for $Q|p_j = p, d_j, \mathcal{M}_j, p-batch, K_i|\sum f_j(T_j)$ and $Q|p_j = p, d_j, \mathcal{M}_j, p-batch, K_i|\max\{f_j(T_j)\}$.

Since $f_j$ ($j = 1, 2, \ldots, n$) is a non-decreasing function, for both $Q|p_j = p, d_j, \mathcal{M}_j, p-batch, K_i|\sum f_j(T_j)$ and $Q|p_j = p, d_j, \mathcal{M}_j, p-batch, K_i|\max\{f_j(T_j)\}$, there is an optimal schedule in which the first batch on each machine starts at time zero, and the batches on each machine are processed successively. Moreover, we can assume that there are $n_i$ empty batches on machine $M_i$ ($i = 1, 2, \ldots, m$) to which the jobs may be assigned, where $n_i$ denotes the smallest integer such that $n_i K_i \geq n$. The $k$-th batch on $M_i$, $B_{k,i}$, completes at time $kp/v_i$, $k = 1, 2, \ldots, n_i$. To find a feasible schedule, we need only to assign the jobs to $\sum_{i=1}^{m} n_i$ empty batches such that all jobs obey the processing set restrictions.

First, we consider $Q|p_j = p, d_j, \mathcal{M}_j, p-batch, K_i|\sum f_j(T_j)$.

If job $j$ is assigned to $B_{k,i}$ and $M_i \in \mathcal{M}_j$, then the cost incurred is defined to be $c_{jki} = f_j(\max\{kp/v_i - d_j, 0\})$, $j = 1, 2, \ldots, n$, $k = 1, 2, \ldots, n_i$, $i = 1, 2, \ldots, m$. Let $C = \max_{j,k,i} c_{jki}$. By regarding each job $j \in \mathcal{J}$ as a vertex in $X$, and each empty batch $B_{k,i}$ as $K_i$ vertices $y_{k1}, y_{k2}, \ldots, y_{kK_i}$ in $Y$, we construct a bipartite graph $G$ with bipartition $(X, Y)$, where $j$ is joined to $y_{k1}, y_{k2}, \ldots, y_{kK_i}$ if and only if $M_i \in \mathcal{M}_j$, and the incurred costs are equal to $c_{jki}$, $j = 1, 2, \ldots, n$, $k = 1, 2, \ldots, n_i$, $i = 1, 2, \ldots, m$. Then we use the Successive Shortest Path algorithm to solve the *bipartite weighted matching problem* [32] and get a matching of minimum cost that saturates every vertex in $X$. From this matching we can construct an optimal schedule easily.

The Successive Shortest Path algorithm runs in $O(|X| \cdot S(|X|+|Y|, |X||Y|, C))$ time, where $S(|X|+|Y|, |X||Y|, C)$ is the time for solving a shortest path problem with $|X|+|Y|$ vertices, $|X||Y|$ edges (these edges have non-negative costs), and maximum coefficient $C$. Currently, $S(u, a, C) = O(a + u \log u)$ [33]. Note that $|X| = n$ and $|Y| \leq 2mn$. We get:

**Theorem 1.** *There is an exact algorithm for $Q | p_j = p, d_j, \mathcal{M}_j, p-batch, K_i | \sum f_j(T_j)$ that runs in $O(n^3 m + n^2 m \log(mn))$ time.*

Next, we consider $Q | p_j = p, d_j, \mathcal{M}_j, p-batch, K_i | \max\{f_j(T_j)\}$. Let $OPT$ denote the objective value of an optimal schedule.

Recall that we are focusing on an optimal schedule in which machine $M_i$ ($i = 1, 2, \ldots, m$) processes $n_i$ batches (some batches may be empty), where $n_i$ denotes the smallest integer such that $n_i K_i \geq n$. Each batch on $M_i$ has $K_i$ positions to accommodate jobs, and there are at most $2n$ positions on $M_i$, $i = 1, 2, \ldots, m$. Therefore, there are at most $2mn$ positions in total. Since each position has at most $n$ choices of accommodating a job, there are at most $2mn^2$ possible values for $OPT$. We can sort these values in ascending order in $O(mn^2 \log(mn))$ time. Then, we perform a binary search in the interval to determine $OPT$ in $O(\log(mn))$ iterations.

For each value $\lambda$ selected, we test whether there is a feasible schedule whose objective value is no more than $\lambda$. To this end, we construct a bipartite graph $G$ with bipartition $(X, Y)$ as follows. Regard each job $j \in \mathcal{J}$ as a vertex in $X$, and each empty batch $B_{k,i}$ as $K_i$ vertices $y_{k1}, y_{k2}, \ldots, y_{kK_i}$ in $Y$, where $j$ is joined to $y_{k1}, y_{k2}, \ldots, y_{kK_i}$ if and only if $M_i \in \mathcal{M}_j$ and $f_j(kp / v_i - d_j) \leq \lambda$, $j = 1, 2, \ldots, n$, $k = 1, 2, \ldots, n_i$, $i = 1, 2, \ldots, m$. Then we use the algorithm in [34] to solve the *maximum cardinality bipartite matching problem*. If the obtained matching saturates every vertex in $X$, then the procedure succeeds and we search the lower half of the interval. Otherwise, the procedure fails and we search the upper half of the interval.

The algorithm in [34] runs in $O(\sqrt{|X|+|Y|} \cdot |X||Y|)$ time. Note that $|X| = n$ and $|Y| \leq 2mn$. We get:

**Theorem 2.** *There is an exact algorithm for $Q | p_j = p, d_j, \mathcal{M}_j, p-batch, K_i | \max\{f_j(T_j)\}$ that runs in $O(n^{5/2} m^{3/2} \log(mn))$ time.*

## 4. Unequal release times

In this section, we consider the case of unequal release times. We will present an algorithm for $Q | r_j, p_j = p, \mathcal{M}_j, p-batch, K_i | C_{\max}$, which generalizes the one in [22] for $Q | r_j, p_j = p, \mathcal{M}_j | C_{\max}$ (the special case of $Q | r_j, p_j = p, \mathcal{M}_j, p-batch, K_i | C_{\max}$ where all $K_i = 1$).

Let $OPT$ denote the makespan of an optimal schedule for $Q | r_j, p_j = p, \mathcal{M}_j, p-batch, K_i | C_{\max}$. Let $\Lambda = \{\lambda | \lambda = r_j + kp / v_i; j, k \in \{1, 2, \ldots n\} \text{ and } i \in \{1, 2, \ldots m\}\}$. The following lemma, adopted from [22], still holds for $Q | r_j, p_j = p, \mathcal{M}_j, p-batch, K_i | C_{\max}$.

**Lemma 1.** $\Lambda$, *as defined above, contains all candidates for $OPT$.*

Since $|\Lambda| \leq mn^2$, there are at most $mn^2$ possible values for $OPT$. We can sort these values in ascending order in $O(mn^2 \log(mn))$ time. Then, we perform a binary search to determine $OPT$ in $O(\log(mn))$ iterations.

For each value $\lambda$ selected, we use the following procedure, AssignJobs, to test whether there is a feasible schedule whose makespan is no more than $\lambda$.

**AssignJobs ($\lambda$):**

Step 1. Assign $b_i = \min\{n_i, \lfloor \lambda \cdot v_i / p \rfloor\}$ empty batches of length $p$ and capacity $K_i$ to machine $M_i$, where $n_i$ denotes the smallest integer such that $n_i K_i \geq n$. The $k$-th batch on $M_i$, $B_{k,i}$, starts at time $\lambda - (b_i - k + 1)p/v_i$ and completes at time $\lambda - (b_i - k)p/v_i$, $k = 1, 2, \ldots, b_i$, $i = 1, 2, \ldots, m$.

Step 2. Construct a bipartite graph $G$ with bipartition $(X, Y)$ as follows. Regard each job $j \in \mathcal{J}$ as a vertex in $X$, and each empty batch $B_{k,i}$ as $K_i$ vertices $y_{k1}, y_{k2}, \ldots, y_{kK_i}$ in $Y$, where $j$ is joined to $y_{k1}, y_{k2}, \ldots, y_{kK_i}$ if and only if $M_i \in \mathcal{M}_j$ and $r_j \leq \lambda - (b_i - k + 1)p/v_i$, $j = 1, 2, \ldots, n$, $k = 1, 2, \ldots, b_i$, $i = 1, 2, \ldots, m$.

Step 3. Use the algorithm in [34] to solve the maximum cardinality bipartite matching problem. If the obtained matching saturates every vertex in $X$, then the procedure succeeds and terminates. Otherwise, the procedure fails and terminates.

**Lemma 2.** *If $OPT \leq \lambda$, then AssignJobs will generate a feasible schedule in $O(n^{5/2} m^{3/2})$ time for $Q \mid r_j, p_j = p, \mathcal{M}_j, p-batch, K_i \mid C_{\max}$ whose makespan is at most $\lambda$.*

*Proof.* Let $\Sigma^*$ denote an optimal schedule. Consider machine $M_i$, $i = 1, 2, \ldots, m$. Since $OPT \leq \lambda$, in $\Sigma^*$, $M_i$ processes at most $b_i$ batches. Without loss of generality, assume that there are $b_i$ batches (some of which may be dummy empty batches) processed on $M_i$ in $\Sigma^*$, denoted as $B_{1,i}^*, B_{2,i}^*, \ldots B_{b_i,i}^*$.

Modify $\Sigma^*$ as follows. Let $B_{b_i,i}^*$ be completed at time $\lambda$, let $B_{b_i-1,i}^*$ be completed at time $\lambda - p/v_i$, .., let $B_{1,i}^*$ be completed at time $\lambda - (b_i - 1)p/v_i$, $i = 1, 2, \ldots, m$. Denote by $\tilde{\Sigma}^*$ the modified schedule. Since each batch in $\tilde{\Sigma}^*$ starts no earlier than its corresponding batch in $\Sigma^*$, $\tilde{\Sigma}^*$ is a feasible schedule. The makespan of $\tilde{\Sigma}^*$ is exactly $\lambda$. Clearly, AssignJobs will generate a feasible schedule which is no worse than $\tilde{\Sigma}^*$.

The algorithm in [34] runs in $O(\sqrt{|X|+|Y|} \cdot |X||Y|)$ time. Since $|X| = n$ and $|Y| \leq 2mn$, the running time of AssignJobs is $O(n^{5/2} m^{3/2})$. □

We get:

**Theorem 3.** *There is an exact algorithm for $Q \mid r_j, p_j = p, \mathcal{M}_j, p-batch, K_i \mid C_{\max}$ that runs in $O(n^{5/2} m^{3/2} \log(mn))$ time.*

## 5. Conclusions

We studied the problem of scheduling jobs with equal lengths and processing set restrictions on uniform parallel batch machines with non-identical capacities. For the case of equal release times, we gave efficient exact algorithms for various objective functions. For the case of unequal release times, we gave efficient exact algorithms for minimizing makespan. The findings extend previous results for uniform machines counterparts. Future research should focus on extending the algorithms to the case of unequal release times for objective functions other than makespan. It would also be interesting to develop algorithms with better time complexities for several special cases of processing set restrictions.

**Competing Interests**

The author confirms that this article content has no competing interests.

**Acknowledgments**

This work is supported by the National Natural Science Foundation of China (nos. 61373079, 61472227, 61672327 and 11501490), Key Project of Chinese Ministry of Education (no. 212101), Shandong Provincial Natural Science Foundation of China (nos. ZR2013FM015 and ZR2015AM006).